\documentclass[proceedings]{JHEP3}

\PrHEP{PrHEP hep2001}                  
\conference{International Europhysics Conference on HEP}                

\usepackage{epsfig}                  

\def\mathswitchr#1{\relax\ifmmode{\mathrm{#1}}\else$\mathrm{#1}$\fi}
\newcommand{\Pf}{\mathswitchr f}

\newcommand{\PW}{\mathswitchr W}

\newcommand{\Pe}{\mathswitchr e}
\newcommand{\Pdbar}{\bar{\mathswitchr d}}
\newcommand{\Pu}{\mathswitchr u}
\newcommand{\TeV}{\unskip\,\mathrm{TeV}}
\newcommand{\GeV}{\unskip\,\mathrm{GeV}}
\newcommand{\MeV}{\unskip\,\mathrm{MeV}}

\renewcommand{\O}{\cal O}

\def\mathswitch#1{\relax\ifmmode#1\else$#1$\fi}

\newcommand{\rw}{\mathswitchr w}

\newcommand{\RacoonWW}{{\sc RacoonWW}} 
\newcommand{\YFSWW}{{\sc YFSWW3}} 

\title{Precise predictions for \boldmath{$\Pe^+\Pe^- \to 4 f (+\gamma)$} 
with anomalous couplings%
\footnote{{\RacoonWW} can be downloaded from 
{\tt http://ltpth.web.psi.ch/racoonww/racoonww.html}.}}

\author{Ansgar Denner\thanks{This work was supported in part by the
    Swiss Bundesamt f\"ur Bildung und Wissenschaft and by the European
    Union under contract HPRN-CT-2000-00149.}
  \\
  Paul Scherrer Institut,
  CH-5232 Villigen PSI, Switzerland\\
  E-mail: \email{Ansgar.Denner@psi.ch}}

\author{Stefan Dittmaier%
        \footnote{Heisenberg Fellow of the 
                  Deutsche Forschungsgemeinschaft DFG}\\            
        Deutsches Elektronen-Synchrotron DESY, 
        D-22603 Hamburg, Germany \\
        E-mail: \email{Stefan.Dittmaier@desy.de}}

\author{\speaker{Markus Roth}\\                   
        Institut f\"ur Theoretische Physik, 
        Universit\"at Karlsruhe,
        D-76128 Karlsruhe, Germany\\                         
        E-mail: \email{roth@particle.uni-karlsruhe.de}}    

\author{Doreen Wackeroth\\            
        Department of Physics and Astronomy, 
        University of Rochester, Rochester, 
        NY~14627-0171, USA \\
        E-mail: \email{dow@pas.rochester.edu}}

\abstract{The relevance of radiative corrections to $4f$
  production at LEP2 and future $\Pe^+ \Pe^-$ colliders is emphasized, 
  and their treatment in the event generator \RacoonWW\ is 
  briefly discussed.
  In particular, the non-universal corrections are compared 
  with the signature of anomalous triple gauge-boson couplings.
  For $4f+\gamma$ production the influence of anomalous
  quartic gauge-boson couplings on photon-energy distributions is 
  illustrated.}

\begin{document}

\section{Relevance of radiative corrections}

From 1996 to 2000, LEP2 has operated very successfully above the
$\PW$-pair threshold allowing for a thorough investigation of
$\PW$-pair production (see e.g.\ Ref.~\cite{lep2} and references
therein).  The results of the four LEP experiments provide accurate
tests of the non-Abelian structure of the Electroweak Standard Model
(SM) and a precise measurement of the $\PW$-boson mass.  While the
precision of the $\PW$-pair production cross section at LEP2 has
reached the per-cent level, the final LEP2 precision for the
$\PW$-boson mass will be 30--$35\MeV$.  At a future $\Pe^+ \Pe^-$
linear collider, the accuracy of the cross-section measurement will be
at the per-mille level, and the precision of $\PW$-mass determination
is expected to be $15 \MeV$ by reconstructing the $\PW$ bosons from
their decay products \cite{MWreconstruct} and about $6 \MeV$ from a
threshold scan \cite{MWthreshold} of the total W-pair cross section.

In order to account for the experimental precision at LEP2, 
%\looseness -1
considerable theoretical effort was undertaken in the past years, as
it is reviewed in Refs.~\cite{Beenakker:1996kt,Grunewald:2000ju}.  In
the present calculations, the $\PW$ bosons are treated as resonances
in the full 4-fermion processes, $\Pe^+ \Pe^- \to 4 \Pf\,
(+\,\gamma)$, and radiative corrections are taken into account in a
proper way.  These can be split into universal and non-universal
corrections.  The former comprise leading-logarithmic corrections (LL)
from initial-state radiation (ISR), higher-order corrections included
upon a proper choice of the input-parameter scheme (running or
effective couplings), and the Coulomb singularity.  The remaining
corrections are called non-universal since they depend on the process
under investigation.  The full $\O(\alpha)$ corrections to the $4f$
processes are not necessary to match the accuracy of LEP2, and it is
sufficient to take only those corrections into account that are
enhanced by two resonant $\PW$~bosons.  The leading term of an
expansion about the two $\PW$ poles provides the so-called {\it
  double-pole approximation} (DPA) \cite{DPA}.  In this approximation,
there are {\it factorizable corrections}, which are the ones to
on-shell $\PW$-pair production and on-shell $\PW$ decay, and {\it
  non-factorizable corrections}, which originate from soft-photon
exchange between the production and decay stages.  Different versions
of such a pole expansion have been used in the literature
\cite{DPAversions,Denner:2000bj,Jadach:2000kw}.  Although several
Monte Carlo programs exist that include universal corrections, only
two event generators, \YFSWW\ \cite{Jadach:2000kw,Jadach:1998hi} and
\RacoonWW\ \cite{Denner:2000bj,Denner:1999gp}, include non-universal
corrections.

While the DPA approach is sufficient for the LEP2 accuracy
\cite{Grunewald:2000ju,Jadach:2001cz}, the extremely high experimental
precision at a future linear collider is a great challenge for future
theoretical predictions, in particular since the DPA is not reliable
near the $\PW$-pair threshold.  The full $\O(\alpha)$ corrections have
to be known for all $4f$ final states, and leading effects beyond
$\O(\alpha)$ have to be included in the theoretical predictions
properly.  For instance, leading and sub-leading Sudakov logarithms
become important and amount to several per cent in the $\TeV$ range
\cite{Denner}.

\section{Non-universal corrections vs.\ anomalous 
triple gauge-boson couplings}

\FIGURE{
\label{fig:ATGC1}
\epsfig{file=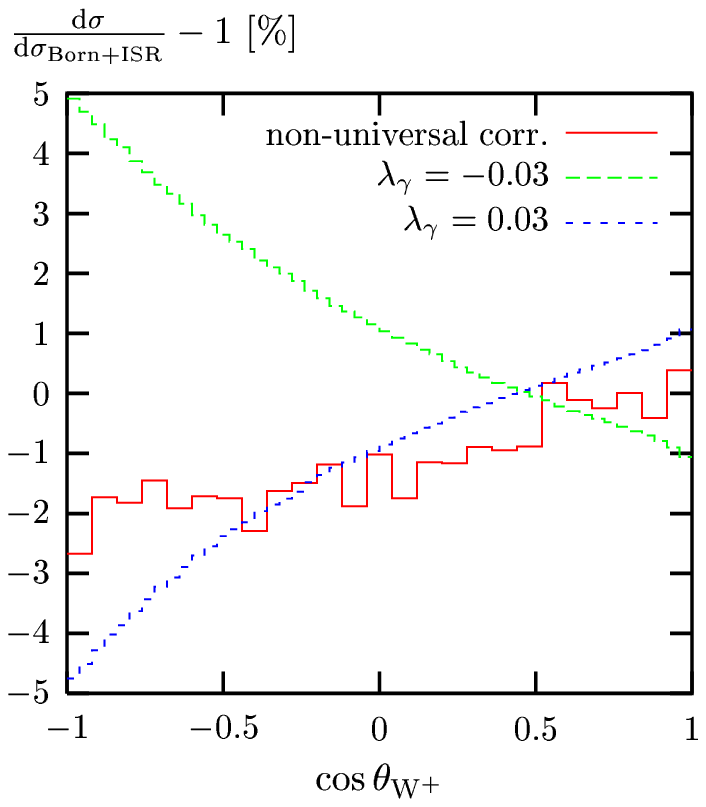,height=8.5cm,width=7cm}
\epsfig{file=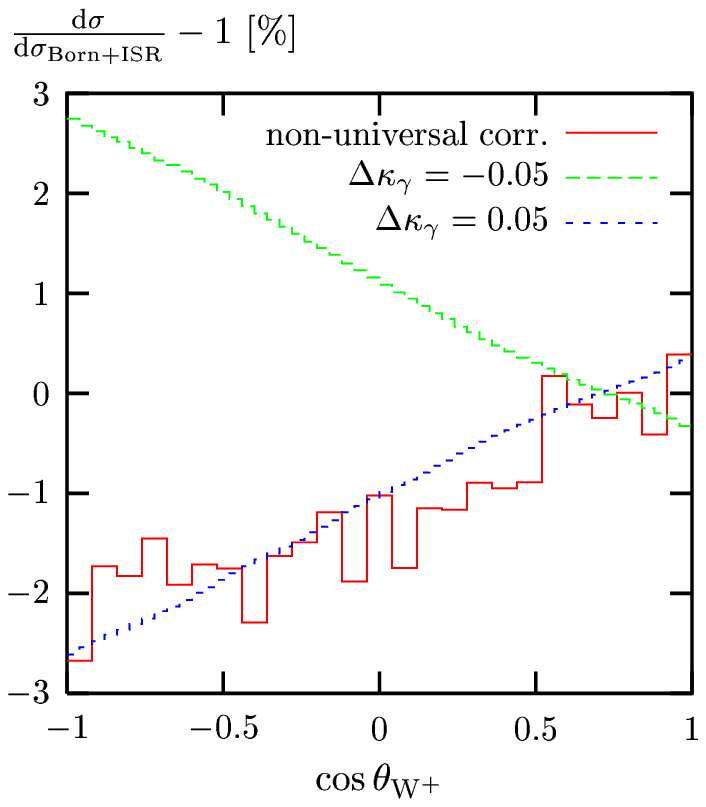,height=8.5cm,width=7cm}
\caption{Influence of anomalous triple gauge-boson couplings and 
  non-universal corrections in the $\PW^+$-production-angle
  distribution for the process $\Pe^+ \Pe^- \to \Pu \Pdbar \mu^-
  \bar\nu_\mu$ at $\sqrt{s}=200 \GeV$ } }% 
As already mentioned, one of the main goals at LEP2 is the
investigation of the non-Abelian structure of the SM.  A proper way to
study possible deviations from the SM is the measurement of anomalous
couplings.  Recently, anomalous quartic \cite{Denner:2001vr} and
triple gauge-boson couplings have been incorporated in \RacoonWW.  A
general parametrization of charged anomalous triple gauge-boson
couplings has been suggested in Ref.~\cite{Hagiwara:1987vm}.  We have
implemented these couplings in \RacoonWW\ in the conventions of
Ref.~\cite{Gounaris:1996rz}, which differ from the former by a sign in
the parity-violating terms.  The implementation of neutral triple
gauge-boson couplings in \RacoonWW\ follows the general
parametrizations of Ref.~\cite{Gounaris:2000kf}.

In the numerical discussion, we compare the influence of the anomalous
charged gauge-boson couplings with the effect of the non-universal
corrections. We adopt the same SM input parameters and the same set of
phase-space cuts as in Refs.~\cite{Grunewald:2000ju,Denner:2000bj}.
In particular, we use the ``bare'' recombination scheme, where the
photon is recombined with a charged fermion if their invariant mass is
smaller than $5 \GeV$.  Following a convention widely used in the LEP2
data analysis, we consider only the P- and C-conserving anomalous
coupling operators with the coupling constants $g^Z_1$, $\kappa_Z$,
$\kappa_\gamma$, $\lambda_Z$, and $\lambda_\gamma$, which are further
constrained by
$$
\Delta \kappa_Z= \Delta g_1^Z - \Delta \kappa_\gamma \tan^2 \theta_\rw , 
\qquad \lambda_Z=\lambda_\gamma, 
$$
where $\theta_\rw$ is the weak mixing angle and $\Delta$ indicates the
deviation from the corresponding SM values
$g_1^Z=\kappa_Z=\kappa_\gamma=1$ and $\lambda_Z=\lambda_\gamma=0$.
Hence, we are left with the three independent parameters $\Delta g^Z_1$,
$\Delta\kappa_\gamma$, and $\lambda_\gamma$.
\FIGURE[h]{
\epsfig{file=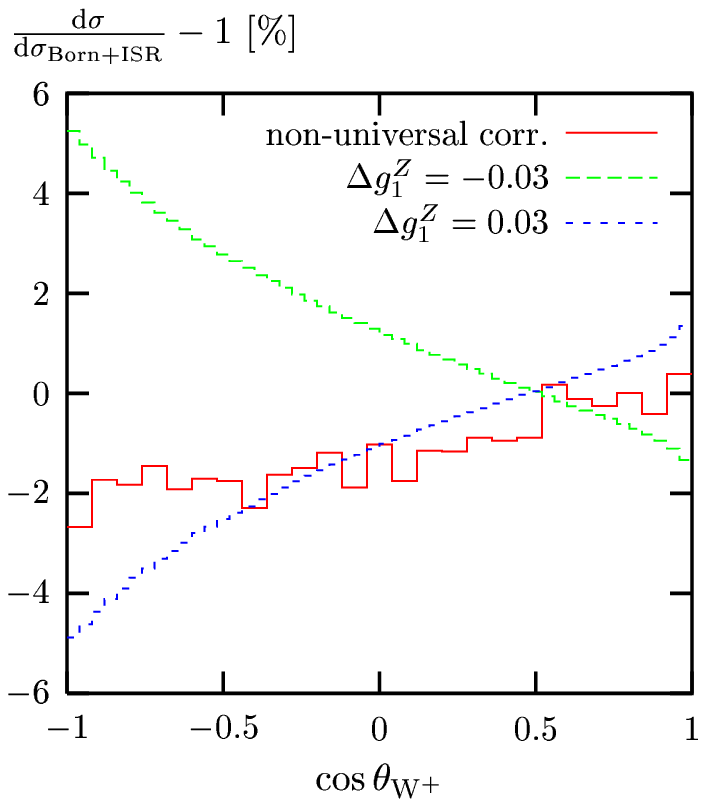,height=8.5cm,width=7cm}
\caption{As in Figure \ref{fig:ATGC1}}
\label{fig:ATGC2}
}%

%\begin{sloppypar}
  In Figures \ref{fig:ATGC1} and \ref{fig:ATGC2} we study the
  influence of the anomalous couplings on the $\PW^+$-production-angle
  distribution for the process $\Pe^+ \Pe^- \to \Pu \Pdbar \mu^-
  \bar\nu_\mu$ at the centre-of-mass energy of $\sqrt{s}=200 \GeV$, as
  predicted by \RacoonWW.  All numbers are normalized to the
  tree-level cross section including higher-order LL ISR up to order
  $O(\alpha^3)$ via structure functions.  The relative deviations for
  different values of the anomalous couplings are compared with the
  corresponding predictions including non-universal $\O(\alpha)$
  corrections instead of anomalous couplings.  The labels indicate the
  values of the corresponding anomalous coupling constants, which are
  chosen to be of the order of the actual accuracy achieved by the LEP
  experiments.  The comparison shows clearly that the non-universal
  corrections are of the same size as the contributions from anomalous
  couplings and, thus, have to be taken into account in the
  determination of the anomalous couplings at LEP2.
%\end{sloppypar}
%\pagebreak[4]

\section{Anomalous quartic gauge-boson couplings in 
  \boldmath{$4f+\gamma$} production} 

Finally, we study the effects of anomalous quartic gauge-boson
couplings in $4f+\gamma$ production. Specifically, we consider the
P-conserving couplings $a_0$, $a_c$, and $a_n$, which have been
investigated in Ref.~\cite{Belanger:1992qh}, and the P-violating
couplings $\tilde a_0$ and $\tilde a_c$, which have been recently
introduced in Ref.~\cite{Denner:2001vr}.  In this analysis we use the
improved Born approximation mode \cite{Denner:2001vr} of \RacoonWW\ 
for $4f + \gamma$ production which includes corrections from
higher-order ISR and from the Coulomb singularity.  We adopt the same
SM input parameters as in Refs.~\cite{Grunewald:2000ju,Denner:2000bj};
in particular, we apply the ADLO/TH cuts.

\FIGURE[b]{
\label{fig:AQGC}
\epsfig{file=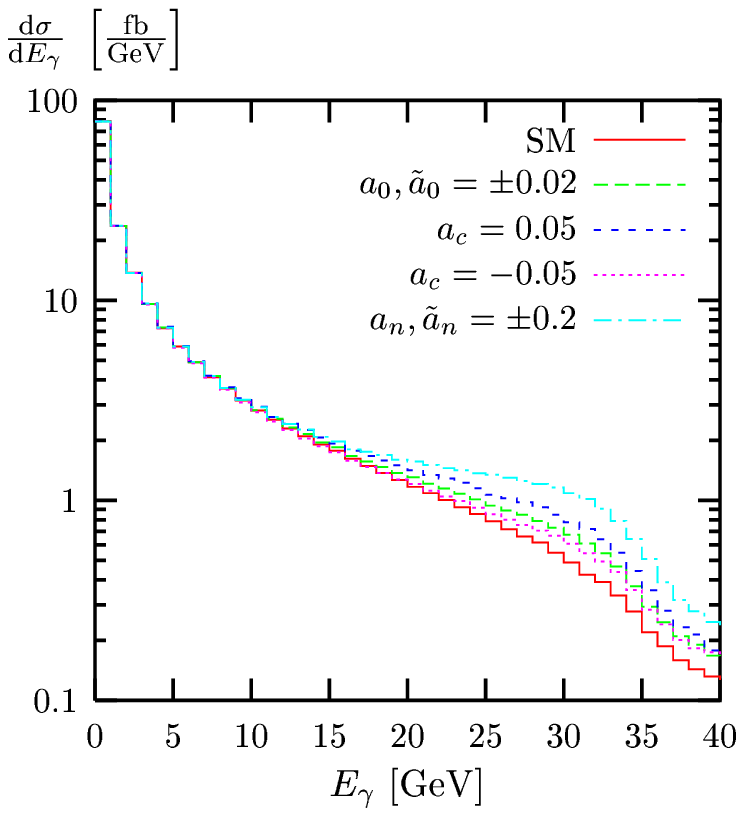,height=8.5cm,width=7cm}
\epsfig{file=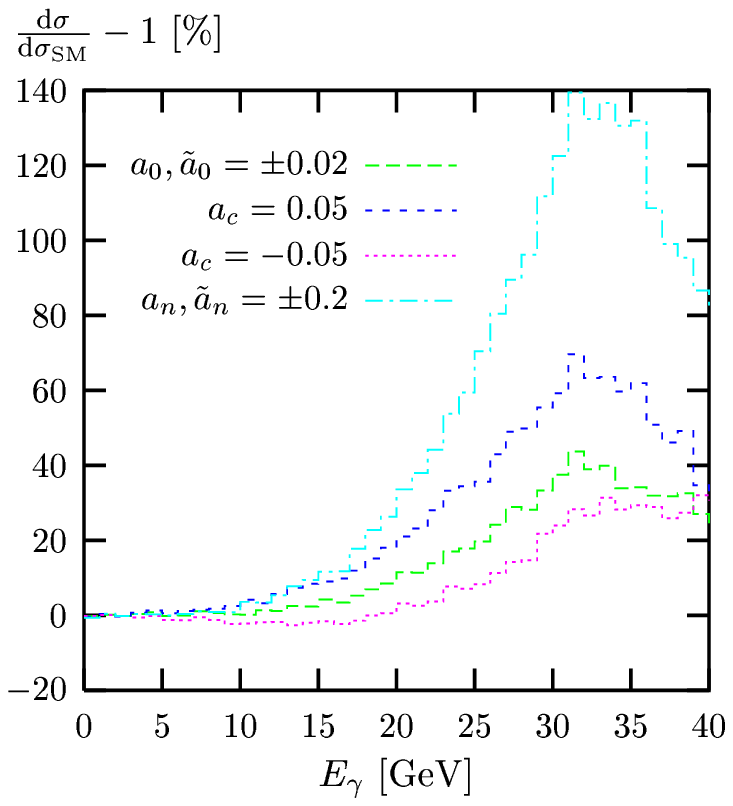,height=8.5cm,width=7cm}
\caption{Influence of anomalous quartic gauge-boson couplings on the 
  photon-energy distribution for the process $\Pe^+ \Pe^- \to \Pu
  \Pdbar \mu^- \bar\nu_\mu \gamma$ at $\sqrt{s}=200 \GeV$ }}
Figure~\ref{fig:AQGC} shows the photon-energy spectrum (l.h.s.)\ and
the corresponding relative deviation from the SM (r.h.s.) for
different values of the anomalous couplings.  The values of the
anomalous couplings are of the order of the present experimental
accuracy.  Only one curve for the coupling $a_0$ and the corresponding
P-violating coupling $\tilde a_0$ (also for $a_n$ and $\tilde a_n$) is
plotted since the respective results are indistinguishable.  Within
the statistical error, the contributions from $a_0$, $a_n$, $\tilde
a_0$, and $\tilde a_n$ are symmetric under a change in sign,
signalling that the quadratic anomalous terms are dominating the
squared amplitude.  However, an effect from changing the sign is
visible for $a_c$, where the interference of anomalous and SM
contributions becomes visible.
%The deviations from the SM are large for hard photons and vanish for 
Since the deviations from the SM are largest for hard photons, effects
from anomalous quartic gauge-boson couplings can be observed best near
the kinematical threshold for \PW-pair production in the $E_\gamma$
distribution.

\end{document}